\documentclass[aps,prl,superscriptaddress,twocolumn]{revtex4}

\usepackage{graphicx} 
\usepackage{amsmath}
 
\newcommand{\beq}{\begin{equation}}
\newcommand{\enq}{\end{equation}}

\begin{document}

\title{Vortex-line solitons in a periodically modulated Bose gas}
\author{J.-P. Martikainen}
\author{H. T. C. Stoof}
\affiliation{Institute for Theoretical Physics, Utrecht University, 
Leuvenlaan 4, 3584 CE Utrecht, The Netherlands}
\date{\today}

\begin{abstract}
We study the nonlinear excitations of a vortex line 
in a Bose-Einstein condensate trapped in a one-dimensional optical lattice.
We find that the classical Euler dynamics of the vortex
results in a description of the vortex line in terms of a
(discrete) one-dimensional Gross-Pitaevskii equation, which allows for both
bright and gray soliton solutions. We discuss these solutions in detail
and predict that it is possible to create vortex-line solitons 
with current experimental capabilities.
\end{abstract}
\maketitle

{\it Introduction.}---
Solitons are special solutions of
nonlinear differential equations with particle-like properties. 
In a Bose-Einstein condensate 
with a repulsive interaction a gray soliton
corresponds to a density minimum for which the loss in interaction energy is
balanced by the increase in the kinetic energy. Gray solitons 
in a Bose-Einstein
condensate have been observed experimentally by imprinting
a $\pi$ phase step onto the condensate 
wave function~\cite{Burger1999a,Denschlag2000a}.
While a Bose-Einstein condensate
with a repulsive interaction is stable against collapse, 
a condensate with attractive interactions may collapse.
This collapse in first instance results 
in ever increasing density gradients. However,
at later stages of the collapse the attractive interaction energy
may be balanced by the kinetic energy, which tends to spread
out the condensate wave function. Therefore, in a one-dimensional condensate
a collapse leads to a creation of a train of bright solitons. Bright solitons 
in a Bose-Einstein condensate have
indeed been observed experimentally~\cite{Strecker2002a,Khaykovich2002a}.
Also theoretically the collapse of the Bose-Einstein condensate
into a train of bright solitons has been 
studied~\cite{Khawaja2002b,Salasnich2002a}.

In this Letter we point out the possibility of vortex-line
solitons in a Bose-Einstein condensate in a one-dimensional
optical lattice. As is well known, the physics of a dilute Bose 
gas in an optical
lattice is described by the Bose-Hubbard 
model~\cite{Jaksch1998a,vanOosten2001a},
which in particular predicts a quantum phase transition from a 
superfluid state into
a Mott-insulator state when the lattice becomes sufficiently
deep. This transition was recently observed by 
Greiner {\it et al.}~\cite{Greiner2002a}.
In the situation of interest to us here the one-dimensional optical lattice 
splits the Bose-Einstein condensate into a stack of weakly-interacting 
pancake condensates, and the vortex line intersects each one
of them. We show that due to the inhomogeneous density distribution 
of the Bose-Einstein condensate, the dynamics of the vortex line is governed
by a (discrete) one-dimensional nonlinear Schr\"{o}dinger or
Gross-Pitaevskii equation with the familiar soliton solutions. 

These solitons are possible when the pressure due to the kinetic energy 
balances the pressure due to the interaction energy. In principle,
these competing processes also exist for 
a vortex line in an ordinary magnetically trapped Bose-Einstein 
condensate without an optical lattice.
In particular, the inhomogeneous condensate-density distribution in the radial
direction gives rise to an interaction-like energy term and the energetic
cost for bending the vortex line gives rise
to the analog of the kinetic energy. However, in an ordinary 
Bose-Einstein condensate the vortex stiffness is orders of magnitude 
larger than in an optical lattice,
where the vortex stiffness can be 
reduced by simply increasing the lattice depth.
In practice this implies that the size of the soliton 
in an ordinary condensate is very large and, in particular, 
larger than the typical condensate size. 
Setting aside other complications this fact alone
suggests that with current experimental capabilities
vortex-line solitons are a possibility that only exists
for a Bose-Einstein condensate in an optical lattice.

So-called optical vortex solitons~\cite{Kivshar1998a} and solitonic
vortices in a Bose-Einstein condensate~\cite{Brand2001a}
have been discussed before. 
Neither one of them is, however, related to the vortex-line
solitons we discuss here. 
The former are related to the non-spreading
propagation of a light beam with a vortex phase pattern 
through a nonlinear medium, while the latter refers to 
the soliton-like properties of a vortex moving through
a very narrow channel. 
Moreover, in classical fluid dynamics Hashimoto~\cite{Hashimoto1972a}
mapped the vortex line dynamics, in the so-called local induction 
approximation, into a nonlinear
Schr\"{o}dinger equation. As a result, solitons predicted by this
formulation are only mathematically 
related to our vortex-line solitons and the physics is fundamentally different.


{\it Theory.}---
We assume a Bose-Einstein condensate
experiencing  a one-dimensional optical lattice potential
in the longitudinal direction. This lattice potential splits the
Bose-Einstein condensate in $N_s$ sites and each site has $N$ atoms.
In the radial direction the atoms are magnetically trapped 
by a harmonic trap with a trapping frequency $\omega_r$. 
The magnetic trapping in the longitudinal direction is assumed to be so 
weak that it can be safely neglected. While the lattice 
is taken to be deep enough to allow us to use a tight-binding approximation 
and to  include only the weak nearest-neighbor Josephson coupling, 
it is also taken to be shallow 
enough to support a superfluid state as opposed 
to the Mott-insulator state~\cite{vanOosten2003a}.

In our earlier work~\cite{Martikainen2003b,Martikainen2004a} we showed that
up to second order in the vortex displacements from the center
of the Bose-Einstein condensate
the Hamiltonian for the vortex line is that of a set of
coupled harmonic oscillators. The eigenmodes of this system
are the Kelvin modes. These modes have been
recently observed in a cigar shaped Bose-Einstein 
condensate in the absence of an optical 
lattice~\cite{Bretin2003a,Mizushima2003a}.
However, if the vortex energy functional is
expanded up to fourth order in the displacements,
the physics of the vortex line turns out to be described 
by a one-dimensional Bose-Hubbard model.
The corresponding Hamiltonian is given by
\begin{eqnarray}
\hat{H}&=&\sum_n\left[\hbar\omega_0+J_V\right]\hat{v}_n^\dagger\hat{v}_n
-\frac{J_V}{2}\sum_{\langle n,m\rangle}\hat{v}_m^\dagger\hat{v}_n
\nonumber\\
&+&\frac{V_0}{2}\sum_n\hat{v}_n^\dagger\hat{v}_n^\dagger\hat{v}_n\hat{v}_n,
\label{eq:H}
\end{eqnarray}
where $\langle n,m\rangle$ denotes the nearest-neighbor layers
and the Kelvon operators $\hat{v}_n$ were defined in terms of the vortex 
displacements $(x_n,y_n)$ as 
$\hat{x}_n=(R/2\sqrt{N})(\hat{v}_n^\dagger+\hat{v}_n)$
and $\hat{y}_n=(iR/2\sqrt{N})(\hat{v}_n^\dagger-\hat{v}_n)$, where
$R$ is the radial size of the pancake condensate.

In view of the complicated dynamics of the three-dimensional
vortex line in a bulk superfluid, 
this relatively simple result is quite remarkable.
Furthermore, as mentioned previously the Bose-Hubbard model shows
a quantum phase transition
from a superfluid into a Mott-insulator state when $J_V/V_0$ becomes
sufficiently small. In our case this transition
describes the break-up of the vortex line into individual pancake vortices
due to quantum fluctuations.
However, in the following we work solely in te superfluid regime, where
a classical mean-field description is appropriate.

Using a Gaussian variational ansatz for the condensate wave function
in each site~\cite{Martikainen2003b}
that is proportional to $\exp\left[-\left(x^2+y^2\right)/2R^2\right]$, we
can analytically calculate the strength of the nearest-neighbor coupling to be
\begin{equation}
J_V=\frac{\hbar\omega_r}{4\pi^2}\Gamma\left[0,\frac{l_r^4}{R^4}\right]
\left(\frac{\omega_L\lambda}{\omega_r l_r}\right)^2
\left(\frac{\pi^2}{4}-1\right)\exp\left(-\frac{\lambda^2 m\omega_L}
{4\hbar}\right)
\nonumber
\label{eq:J}
\end{equation}
and the interaction strength
\beq
V_0=\frac{2\hbar\omega_r\left(l_r/R\right)^2\Gamma\left[0,l_r^4/R^4\right]-3\hbar\omega_r
\left(l_r/R\right)^2-4\hbar\Omega}{4N}\nonumber,
\label{eq:V0}
\enq
where $\omega_L=\sqrt{8\pi^2V_L/m\lambda^2}$ is the 
oscillator frequency of the optical lattice, $l_r=\sqrt{\hbar/m\omega_r}$,
$\Gamma\left[a,z\right]$ is the incomplete gamma function,
$V_L\cos^2\left(2\pi z/\lambda\right)$ is the lattice potential,
$\lambda$ is the wavelength of the laser light,
and $\Omega$ is the rotation frequency of the magnetic trap.
Furthermore, a straight and slightly displaced vortex precesses around 
the condensate center with 
the frequency $\omega_0=\left(\omega_r l_r^2/2R^2\right)
\left(1-\Gamma\left[0,l_r^4/R^4\right]\right)+\Omega$.
The radial size $R$ of the Bose-Einstein condensate is determined by
minimizing the condensate energy with the vortex line in the center of the
condensate~\cite{Martikainen2004a}.

In Ref.~\cite{Martikainen2003d} we discussed the equilibrium 
properties of a straight vortex when the interaction term is
attractive and pointed out that in this case the quantum mechanical
state of the vortex  can become strongly squeezed. 
However, here make the usual classical approximation
and study the nonlinear excitations of the vortex line. The solitonic
solutions that we find can in principle also be squeezed under
the appropriate conditions, but this interesting topic is deferred to
future work.


Under the assumption of a coherent state for the Bose-Einstein
condensate, the Euler dynamics of the vortex line 
in an optical lattice is given by the discrete 
Gross-Pitaevskii equation
\begin{eqnarray}
i\dot{v}_n(t)&=&-\frac{J_V}{2}\left[v_{n-1}(t)-2v_n(t)+v_{n+1}(t)\right]
\nonumber\\
&+&\left[\omega_0+V_0|v_n(t)|^2\right]v_n(t),
\label{eq:vgp}
\end{eqnarray}
where from now on we use units such that $\hbar=1$.
If the variation of the vortex positions is small over one lattice spacing
this equation is simply the discretized version of the continuum
Gross-Pitaevskii equation
\beq
i\dot{v}_n(t)=-\frac{J_V}{2}\frac{\partial^2 v_n(t)}{\partial n^2}+
\left[\omega_0+V_0|v_n(t)|^2\right]v_n(t).
\label{eq:GP}
\enq
As indicated by the above equation we from now on
also choose to work with 
a longitudinal unit of length that equals the lattice spacing $\lambda/2$. 

{\it Dark vortex-line soliton.}--- 
Dark or gray soliton solutions of Eq.~(\ref{eq:GP}) are possible when
the interaction is repulsive, {\rm i.e.}, $V_0>0$. 
This is the case in the non-rotating
trap, for example. Explicitly the gray soliton solution moving with
velocity $v$ is given by
\beq
v_n(t)=\sqrt{n_0}\left\{i\cos \theta+
\sin\theta\tanh\left[\sin \theta\frac{(n-v t)}{\xi}\right]\right\}e^{-i\mu t},
\nonumber
\enq  
where $n_0=\lim_{n\rightarrow\infty} |v_n|^2$, $\cos \theta=v/c$, 
$c=\sqrt{J_VV_0n_0}$ is the speed of sound,
$\mu=\omega_0+V_0n_0$, and the soliton size
is $\xi=\sqrt{J_V/V_0n_0}$ lattice spacings.
While the gray soliton solution is simple to construct theoretically,
its experimental preparation is challenging since it
requires a good control over the position of the vortex in each site. 
One possible way to create a dark soliton
is to prepare a vortex line that is tilted 
with respect to the optical potential.
If the vortex line crosses the center of the condensate
somewhere along the lattice, this initial state provides a prototype
for the dark soliton. Other initial states can similarly
resemble gray solitons.
Unfortunately, the ensuing time evolution, which we have studied
numerically,
typically also generates a fair number of other vortex-line
excitations, such as Kelvin modes and phonons. This makes
the experimental observation complicated.

\begin{figure}
\includegraphics[width=1.0\columnwidth]{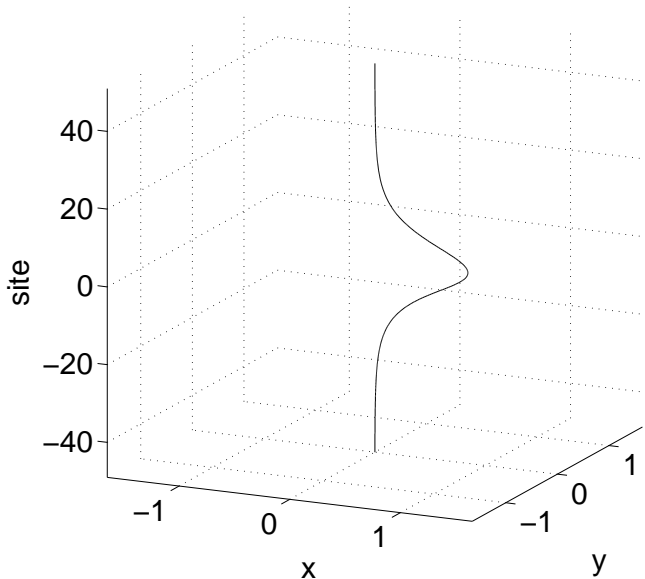}
\vspace{-0.5cm}
\caption[Fig1]{A bright soliton solution of Eq.~(\ref{eq:brightsoliton})
with $N_0=1000$, $k=0$, $J_V=0.4\,\hbar\omega_r$, 
and $V_0=-7\cdot 10^{-5}\,\hbar\omega_r$.
These values for $J_V$ and $V_0$ are realistic in an optical potential with
a lattice spacing $d=398\, {\rm nm}$, depth $V_L=15\, E_r$, with
$1000$ $^{87}{\rm Rb}$ atoms in each site, and a rotation frequency
$\Omega=0.2\,\omega_r$. here $\omega_r=2\pi\cdot 100\,{\rm Hz}$ is the radial
trapping frequency, $E_r=h^2/2m\lambda^2$ is the recoil energy,
and the unit of length in the transverse direction is 
$\sqrt{\hbar/m\omega_r}$.}
\label{fig:brightsoliton}
\end{figure}

{\it Bright vortex-line soliton.}--- Since experimental 
preparation of the dark soliton is 
complicated, we focus now on the regime where the interaction
is attractive, {\rm i.e}, $V_0<0$. 
This occurs, when the magnetic trap is rotated so fast that
the vortex is energetically stable.
With an attractive interaction there exists a solution 
of Eq.~(\ref{eq:GP}) corresponding
to the bright vortex-line soliton moving along the lattice,
which is given by
\beq
v_n(t)=\sqrt{\frac{N_0}{2\xi}}\frac{e^{-i\mu t+ikn}}
{\cosh\left[\left(n-J_V k t/2\right)/\xi\right]},
\label{eq:brightsoliton}
\enq
where $k$ is the wavevector, $\mu=\omega_0-J_V/2\xi^2+J_V k^2/2$,
and the size of the
soliton is $\xi=2J_V/|V_0|N$ lattice spacings. This solution is
normalized such that $\int dn |v_n|^2=N_0$. 
When $k=0$, this solution is equivalent
with a shape preserving precession 
of the curved vortex line. Since
the chemical potential $\mu$ of the stationary soliton
is smaller than $\omega_0$,
the soliton precesses around the center
of the condensate with a lower precession frequency than
a straight vortex line. In Fig.~\ref{fig:brightsoliton}
we show an example of the bright vortex soliton with 
typical parameter values.

\begin{figure}
\includegraphics[width=0.9\columnwidth]{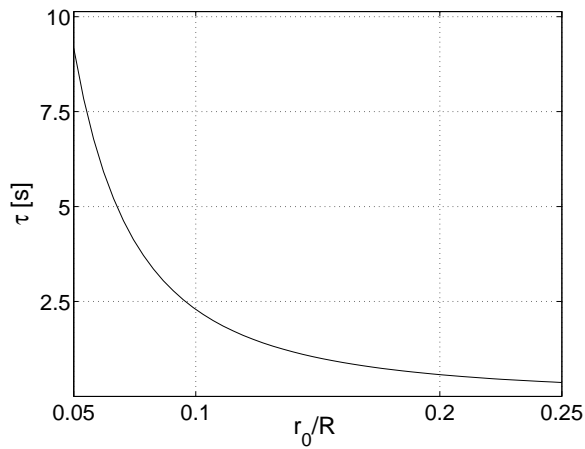}
\vspace{-0.3cm}
\caption[Fig2]{The timescale for the vortex line collapse
as a function of the initial displacement $r_0$
relative to the condensate size $R$. For concreteness we further
assumed that
$1000$ $^{87}{\rm Rb}$ atoms were trapped in each site with 
$\Omega=0.2\,\omega_r$ and $\omega_r=2\pi\cdot 100\,{\rm Hz}$. 
The lattice depth was $V_L=15\, E_r$ and the lattice 
spacing was $d=398\, {\rm nm}$.
}
\label{fig:timescale}
\end{figure}

{\it Soliton creation.}--- 
In order to gain quantitative understanding of the vortex-line collapse
into a train of such solitons, we need to study the excitations around
the homogeneous solution.
At long wavelengths the Bogoliubov dispersion for a displaced but
straight infinitely long vortex line corresponding to the density 
$|v_n|^2=n_0$
is given by
\beq
\epsilon_k=J_V\sqrt{\frac{k^2}{2}\left(\frac{k^2}{2}
+\frac{2V_0n_0}{J_V}\right)}.
\label{eq:bogdispersion}
\enq
At wavelengths that are comparable to the lattice spacing the 
lattice discreteness becomes apparent and the dispersion is modified
by the replacement $J_Vk^2/2\rightarrow J_V\left[1-\cos(k)\right]$.
However, for our purposes the long wavelength limit of the dispersion
is sufficient. 
When $V_0<0$ the dispersion has imaginary energies and the
system is dynamically unstable.
The most unstable mode has the wavevector $k_0=\sqrt{2|V_0|n_0/J_V}$
and the absolute value of its energy can be used to estimate
the characteristic timescale $\tau$ for the collapse by
$
\tau=1/|\epsilon_{k_0}|=1/|V_0|n_0.
$
In contrast to the nearest-neighbor coupling strength $J_V$, $V_0$ 
is not very sensitive to the depth of the lattice potential.
Therefore, the timescale for the vortex-line collapse 
is hard to tune by changing the lattice depth. A better way
to tune the collapse time is to vary the initial vortex displacement.
We demonstrate this in Fig.~\ref{fig:timescale}, which 
shows the behavior of $\tau$ with some
typical parameter values as a function of the initial vortex-line
displacement. It is clear from this figure that by displacing the
vortex line sufficiently, the collapse can 
occur so quickly that the dissipative processes are unlikely to
affect the vortex dynamics.

As the wavevector $k_0=2\pi/\lambda_0$ indicates 
the wavelength $\lambda_0$
of the most unstable vortex-line fluctuations, we 
can estimate that the collapse creates 
about $N_s k_0/2\pi$ solitons.
This in turn can be used to estimate 
the size of the solitons after the collapse. We find, using
the lattice spacing as the unit of length,
that the soliton size after collapse is
given by $\xi=\sqrt{2J_V/|V_0|n_0}/\pi$. 
However, it is clear that the above arguments
are only valid if the final soliton size is much smaller than
the system size $N_s$. Otherwise, boundary effects are expected
to influence the results considerably. 
Another way to understand this is to
note that in the finite lattice the smallest nonzero
wave vector is given by $k_{min}=2\pi/N_s$.
Moreover, we can see from Eq.~(\ref{eq:bogdispersion})
that the dynamical instability exist only in the regime
where $|k|<2\sqrt{|V_0|n_0/J_V}=k_c$.
Therefore, the vortex line in an optical lattice can only be
dynamically unstable if $k_{min}\ll k_c$. This implies the condition
\beq
\frac{\pi}{N_s}\sqrt{\frac{J_V}{|V_0|n_0}}\ll 1,
\label{eq:collapsecriteria}
\enq
which turns out to be equivalent to the requirement that the final soliton
size must be much smaller than the system size.

\begin{figure}
\includegraphics[width=\columnwidth]{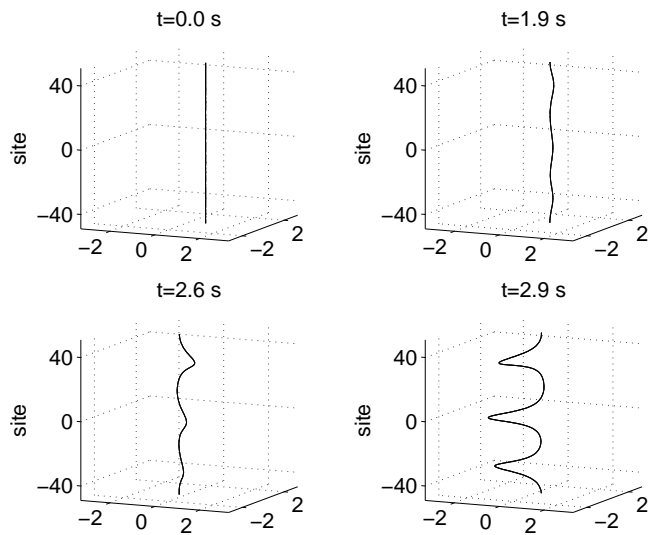}
\vspace{-0.5cm}
\caption[Fig3]{Time evolution of the vortex line 
during its collapse into a train of three bright solitons.
Initially the vortex line was displaced from 
the center of the Bose-Einstein condensate by $R/4$. Furthermore,
the time evolution was seeded by an unobservable amount of white noise
to simulate experimental fluctuations.
Each site had 
$1000$ $^{87}{\rm Rb}$ atoms, the lattice depth was $V_L=15\, E_r$,
$\Omega=0.2\,\omega_r$, and $\omega_r=2\pi\cdot 100\,{\rm Hz}$. 
The lattice spacing was $d=398\, {\rm nm}$.
The transverse unit of length is $\sqrt{\hbar/m\omega_r}$.
}
\label{fig:vortexcollapse}
\end{figure}

The above arguments based on a Bogoliubov approach are valid only when
the deviations from the initial homogeneous system are small.
Given enough time the disturbances in the dynamically unstable system
grow large and the Bogoliubov approach fails. In this limit
we have to solve the discrete time-dependent Gross-Pitaevskii equation
numerically. In Fig.~\ref{fig:vortexcollapse} we show the time
evolution of a displaced vortex line that was initially almost straight.
With the parameters used in this figure, which was obtained
by a fourth-order Runger-Kutta method, the vortex line is seen to
collapse into three bright solitons.

{\it Discussion.}---
The {\it in situ} imagining of 
the three-dimensional vortex line is difficult, but not
impossible~\cite{Bretin2003a,Anderson2001a}. However,
in many experiments the magnetic trap potential is turned 
off and the condensate
is allowed to expand in order to facilitate the imaging of the
Bose-Einstein condensate. In an earlier 
paper~\cite{Martikainen2004a} we showed
that during the expansion the structure of the three-dimensional vortex line
in a good approximation only changes by 
a scale factor and therefore vortex-line solitons are 
observable even after the expansion of the Bose-Einstein condensate.

This offers the opportunity to study experimentally 
a variety of other interesting phenomena.
One intriguing possibility is to study a Bose-Einstein condensate
with more than one vortex line. In this manner it is possible to
study how solitons in different vortex lines interact. 
The possibility of breather solutions is also 
an exiting prospect~\cite{Trombettoni2001b}.
 
We thank D. van Oosten for helpful remarks.
This work is supported by the Stichting voor Fundamenteel Onderzoek der 
Materie (FOM) and by the Nederlandse Organisatie voor 
Wetenschaplijk Onderzoek (NWO).

\end{document}